# Robust local empirical Bayes correction for Bayesian modeling

Yoshiko Hayashi[1]

Osaka Central Advanced Mathematical Institute

Osaka Metropolitan University

## Abstract

This paper investigates a robust empirical Bayes correction for Bayesian modeling. We show the application of the model on income distribution. Income shock includes temporal and permanent shocks. We aim to eliminate temporal shock and permanent shock using two-step local empirical correction method. Our results show that only 6.7% of the observed income shocks were permanent shock, and the posterior (permanent) mean weekly income was reduced from the observed income ₤415 to ₤202 for the United Kingdom using the Living Costs and Food Survey in 2021-2022.

[1] yoshiko-hayashi@omu.ac.jp
3-3-138 Sugimoto Sumiyoshi Osaka 558-8585

## 1. Introduction

Efron (2011) proposed an empirical local Bayes correction that allows a more general form than the James-Stein estimation. Hayashi (2025) extended the local Bayesian correction to Bayesian modeling. The model is a two-stage empirical local Bayes correction model, that does not require to specify the form of the prior distribution.

Furthermore, we develop a robust model for our study. Bayesian robustness modelling using Studen-$t$ distribution, provides a theoretical solution to the outlier problem. Dawid (1973) formally provided a theoretical solution to the conflict of information, i.e., the conflict between prior distribution and data. Andrade and O'Hagan (2011) provided sufficient conditions for robust modelling using regular variation theory. O'Hagan and Pericchi (2012) reviewed previous studies that used this model. Gagnon and Hayashi (2022) presented a theoretical analysis of the Studen-$t$ linear regression model. We used Student-$t$ distribution for robust modelling based on these studies. We apply the model to income distribution data.

This paper organised as follows. Section 2 provides an overview of the local empirical Bayes correction and the extension based on Hayashi (2025). Section 3 presents the application results for the United Kingdom using local empirical Bayes correction. Finally, Section 4 discusses the results and future work.

## 2. Empirical Bayes correction to the Bayesian model

### 2.1 Local empirical Bayes correction

Efron (2011) provides a local empirical Bayes correction that allows multiple modes for the prior distribution of the location parameter. The Tweedie distribution-based local empirical Bayes correction is expressed as follows:

$$\mu|y \sim \left(y + \sigma^2 l'(y), \sigma^2 \left(1 + \sigma^2 l''(y)\right)\right), \qquad (1)$$

where $l(y)$ represents the logarithm of the likelihood, thus, the posterior mean, $\widehat{\mu_i}$ becomes observational information with the Bayesian corrected term.

$$\widehat{\mu_i} \equiv \hat{E}(\mu_i|y_i) = y_i + \sigma^2 \hat{l}'(y_i). \qquad (2)$$

To estimate $l(y)$, Efron (2011) used Lindsey's method, in which the frequency divided into $K$ bins follows an independent Poisson distribution. We adopted this estimate and obtained the marginal distribution from the $J$-th order Poisson regression.

### 2.2. Application of local empirical Bayes correction using the median

This study applied Hayashi's (2025) modification to income distribution data. Efron's (2011) local empirical Bayes correction requires that the distribution of each observation follow an exponential distribution. To address this problem, we used Efron's (2011) conversion scores.

$$z \sim \Phi^{-1}(F(x)), \qquad (3)$$

where $\Phi$ and $F(x)$ denote the distribution functions of the standard normal distribution and observation, respectively. Thus, the transformation can be applied to any distribution, including Student-$t$ distribution. This transformation was used to obtain scores that followed a standard normal distribution.

Ventrucci and Scott (2011) considered the posterior probability that the mean posterior is larger than some value, $m$. The mean posterior is determined as follows:

$$P_{M_i} \equiv Pr\{\mu_i > m|data\}. \qquad (4)$$

Consider the posterior probability of the following:

$$P_{M_i}^* \equiv Pr\{\mu_i > \mu_i^* | data\}, \quad (5)$$

if $P_{M_i}^* = 0.50$, $\mu_i^*$ becomes the median. Thus, if we use $\mu_i^*$ as our score, we can treat it as the mean of the $z$ score from Equation (3). This confirms that the Tweedie formula is applicable.

To estimate the standard deviation, Efron and Zhang (2011) used half of the distance between the 16th and 84th percentiles as the robust standard deviation, which is denoted as $S_R$. Therefore, we also used the $S_R$ from the $\mu_i^*$ distribution.

## 3. UK income distribution analysis

It is popular procedure that decomposes residual variation of income into temporary and permanent components defined as residuals and a random-walk permanent component, using time series data.

In our model, the temporal and permanent income shocks are defined as follows. Permanent income is the expected lifetime income based on all information at the time. Thus, we define the permanent income as the posterior mean, $\mu_i|y_i$, and the temporal income as the divergence from the observed income, $y_i - \mu_i|y_i$. This paper estimates and corrects the posterior means using local empirical Bayes correction.

### 3.1 Data set

We estimated the income distribution density in the UK. The data comes from the Living Costs and Food Survey (LCF) 2021-2022, a continuous survey of samples of the household-dwelling UK population. The data were provided by the Data Archive at the University of Essex, Department of Employment, Statistics Division. We used disposable personal income (i.e., after-tax and transfer income). Our definition of the unemployed excludes job-searching unemployed people (see flowcharts of LCF derived variables). The number of observations was $n$= 10,504.

Figure 1 shows the income distribution histogram and kernel density. Table 1 shows the basic income data statistics.

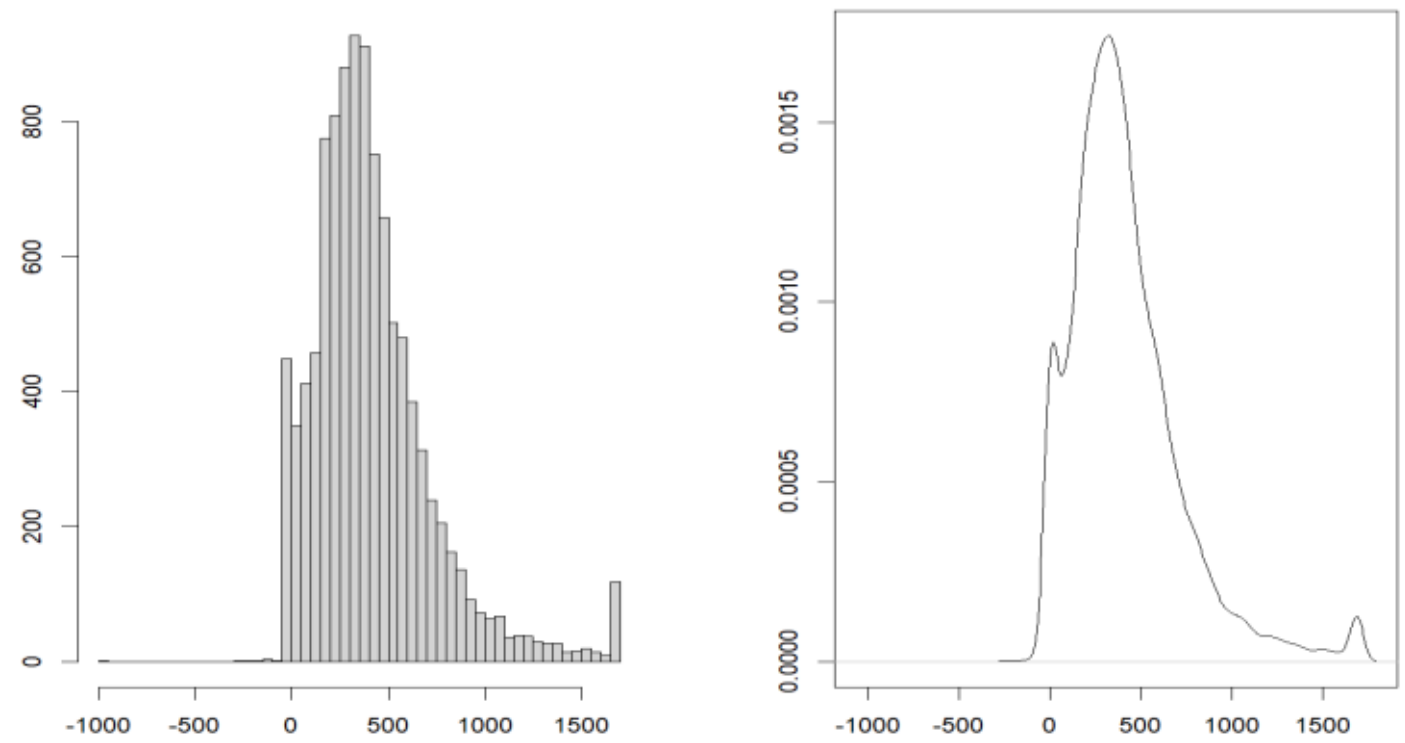

**Figure 1**: Histogram and Kernel density of original data

**Table 1: Basic income statistics**

| Minimum | 1st Quartile | Mean | 3rd Quartile | Maximum |
|---|---|---|---|---|
| -965.6 | -211.7 | 414.9 | 548.4 | 1691.6 |

### 3.2 Model

The model was constructed in two steps. First, we obtained the posterior $z$ score using hierarchical modelling. Second, we applied Efron's (2011) local empirical Bayes correction.

#### 3.2.1 First Step

To obtain the median of the posterior distribution for each observation, we used Student-$t$ distribution with four degrees of freedom for likelihood. Therefore, an observation was disregarded if it was located far from the mean. The priors for the location and scale parameters followed the normal distribution and the Jeffreys prior, which was introduced by Fonseca et al. (2008), respectively. The hyperparameters follow a normal distribution with a mean of 0 and variance of 1,000,000 for the location hyperparameter and a Gamma distribution with a mean of one and a variance of 100 for the scale hyperparameter.

$$\begin{cases} f(y \mid \mu,\sigma) = t_4(\mu,\sigma) \\ \mu \overset{D}{\sim} N(\mu_0,\sigma_0) \\ \log(\sigma) \overset{D}{\sim} Uniform \\ \mu_0 \overset{D}{\sim} N(0,1000000) \\ \sigma_0 \overset{D}{\sim} Gamma(0.01,0.01) \end{cases} \tag{6}$$

We created the model using OpenBUGS(v.3.2.3) through R(v.4.2.3) to obtain the posterior distribution. The first 1,000 samples were removed during the 11,000 iterations. To estimate the distribution, we adopted Lindsey's method, which was also used by Efron (2011) with the glm package in R(v.4.2.3). The data are standardized as $x_i^* = \frac{\mu_i^* - \bar{\mu}}{s_R}$ and the basic statistics are shown in Table 2. To generate the histogram, we used a width of 0.25 for the standardized income data interval ranging from -2.00 to 4.50.

**Table 2: Basic statistics for standardized income**

| Minimum | 1st Quartile | Mean | 3rd Quartile | Maximum | sd |
|---|---|---|---|---|---|
| -4.460 | -0.656 | 0 | 0.431 | 4.124 | 1 |

**Table 3: Basic statistics for the posterior *z* score of income**

| Minimum | 1st Quartile | Mean | 3rd Quartile | Maximum | sd |
|---|---|---|---|---|---|
| -0.637 | -0.242 | -0.170 | -0.094 | 0.302 | 0.117 |

Table 3 shows the basic statistics for non-corrected posterior m eans. Figure 2 shows that the posterior distribution locates lower th an the original, and the extremes have been eliminated as expected.The standard deviation decreases from 1 to 0.117 in this stage.

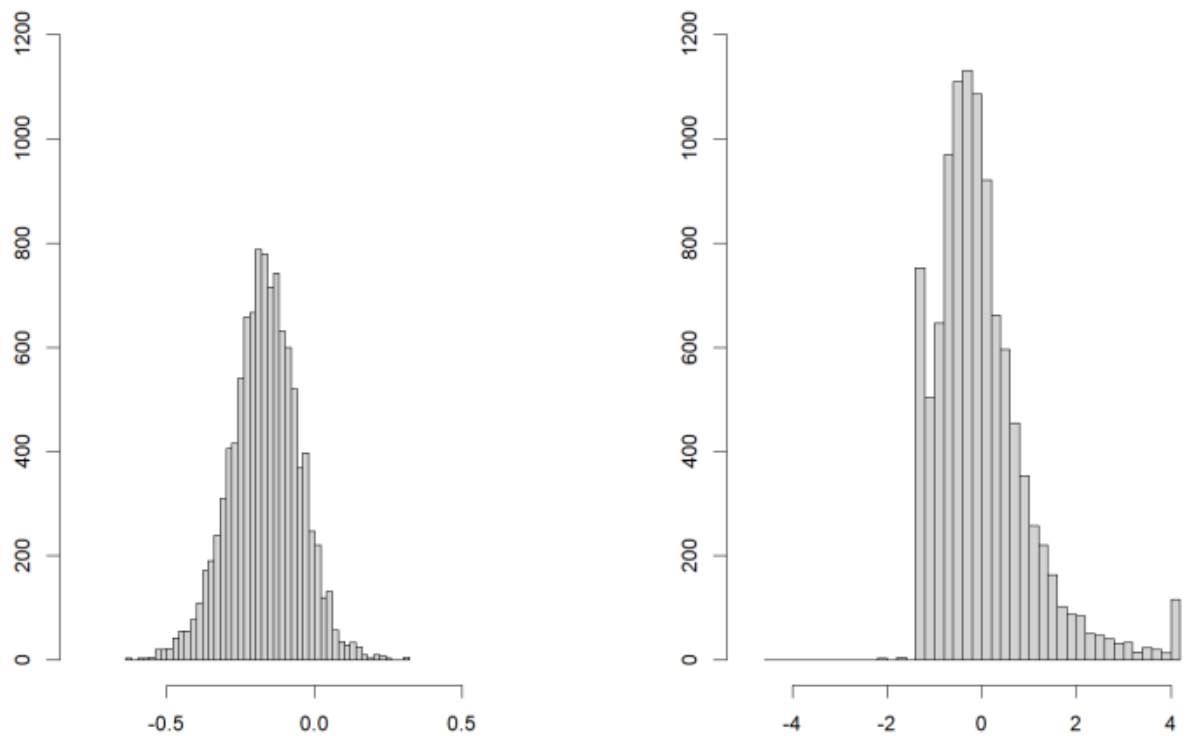

**Figure 2:** The left figure shows the histogram of the posterior *z* score, and the right figure represents the original *z* score.

### 3.2.2 Second Step

In this section, we apply Efron's (2011) local empirical Bayes correction to the posterior *z* scores obtained in the first step. We created 100 bins for the histogram and determined the order of the Poisson regression using AIC. Table 4 shows the AIC results from the 2nd to 7th orders. Based on the AIC results, we selected the $5^{th}$-order model for Poisson regression.

**Table 4: AIC of Poisson regression**

| # of order | 2 | 3 | 4 | 5 | 6 | 7 |
|---|---|---|---|---|---|---|
| AIC | 870.01 | 860.24 | 764.92 | 749.40 | 749.41 | 750.94 |

**Table 5: Basic statistics for the corrected posterior z score of income**

| Minimum | $1^{st}$ Quartile | Mean | $3^{rd}$ Quartile | Maximum | sd |
|---|---|---|---|---|---|
| -0.773 | -0.290 | -0.259 | -0.222 | 0.134 | 0.067 |

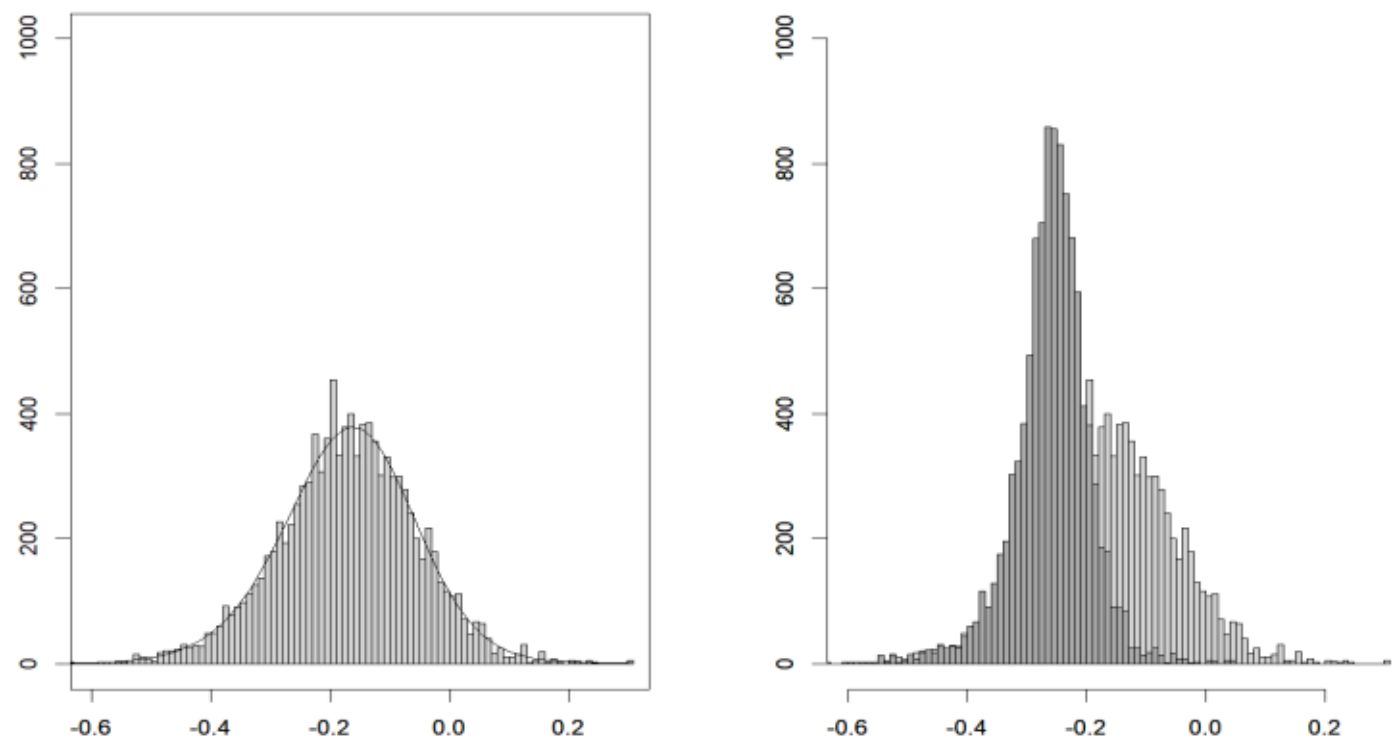

**Figure 3:** The left figure shows the fitted line of Poisson regression, and the right figure shows the original (light gray) and corrected (dark gray) distributions.

Table 5 presents the basic statistics of the corrected posterior *z* scores. The left panel of Figure 3 shows a histogram of the median score and the fitted line of the 5th-order Poisson regression. The right panel shows histograms of the original standardized median income and corrected distributions.

The mean and standard deviation of the corrected posterior *z* s cores were -0.259 and 0.067, respectively. Thus, the corrected mean in come of the original scale becomes ₤202. the standard deviation was re duced more by 42.4% than non-corrected posterior *z* score. Since the ori ginal standard deviation of *z* score which includes both temporary and permanent components is 1 and the corrected one is 0.067, only 6.7% o f income shock is the permanent income shock.

## 4. Summary

This study investigates a two-step local empirical Bayes correction method. Our results show that only 6.7% of income shocks observed are temporary using a local empirical Bayes correction. The UK income distribution's corrected mean weekly income was determined to be ₤202.